\documentclass{ws-ijmpe}

\begin{document}

\markboth{R.\ Alkofer, C.S.\ Fischer, F.\ Llanes-Estrada, K.\ Schwenzer}{
What the Infrared Behaviour of QCD Vertex Functions in  Landau gauge
can tell us about Confinement}

\catchline{}{}{}{}{}
%%%%%%%%%%%%%%%%%%%%%%%%%%%%%%%%%%%%%%%%%%%%%%%%%%%%%%%%%%%%%%%%%%%%

\title{What the Infrared Behaviour of QCD Vertex Functions\\ in  Landau gauge
can tell us about Confinement}

\author{Reinhard Alkofer}

\address{Department of Theoretical Physics, Graz University,
Universit\"atsplatz 5, A-8010 Graz, Austria\\
reinhard.alkofer@uni-graz.at}

\author{Christian S. Fischer}

\address{Institut f\"ur Kernphysik, TU Darmstadt, Schlossgartenstrasse 9, 64289
Darmstadt, Germany \\
christian.fischer@physik.tu-darmstadt.de}

\author{Felipe J.\ Llanes-Estrada}

\address{Dept. F\'{\i}sica Teorica I, Univ. Complutense, Madrid 28040, Spain\\
fllanes@fis.ucm.es}

\author{Kai Schwenzer}

\address{Department of Theoretical Physics, Graz University,
Universit\"atsplatz 5, A-8010 Graz, Austria\\
kai.schwenzer@uni-graz.at}

\maketitle

\begin{history}
\received{(received date)}
\revised{(revised date)}
%\accepted{(Day Month Year)}
%\comby{(xxxxxxxxxx)}
\end{history}

\begin{abstract}
The infrared behaviour of Landau gauge QCD vertex functions is investigated
employing a skeleton expansion of the Dyson-Schwinger and Renormalization Group
equations. Results for the ghost-gluon, three-gluon, four-gluon and quark-gluon
vertex functions are presented. Positivity violation of the gluon propagator,
and thus gluon confinement, is demonstrated. Results of the Dyson-Schwinger
equations for a finite volume are compared to corresponding lattice data.
It is analytically demonstrated that a linear rising potential between heavy
quarks can be generated by infrared singularities in the dressed  quark-gluon
vertex. The selfconsistent mechanism that generates these singularities
necessarily entails the scalar Dirac amplitudes of the full vertex and the quark
propagator. These can only be present when chiral symmetry is broken, either
explicitly or dynamically. 
\end{abstract}

\section{On theories of Quark Confinement}

Quark confinement is definitely the hardest problem in hadron physics. Over the
last three decades many theories have been suggested to elucidate this
phenomenon. It has turned out that the main challenge for such theories is
posed by the properties of the linearly rising static quark-antiquark potential
as uncovered by many Monte-Carlo lattice calculations, see {\it e.g.} ref.\
\cite{Alkofer:2006fu} and references therein for a corresponding discussion.

Theories of confinement which are currently debated include ones based on
\begin{romanlist}[(ii)]
\item the condensation of chromomagnetic monopoles \cite{Mandelstam,Pisa},
\item the percolation of center vortices \cite{Greensite},
\item the Gribov-Zwanziger scenario in Coulomb gauge \cite{Gribov,Dan1}
\item the infrared behaviour of Landau gauge Greens functions
\cite{Alkofer:2000wg,Fischer:2006ub,Aguilar:2004sw}, and
\item the AdS$_5$ / QCD correspondence \cite{Maldacena:1998im}.
\end{romanlist} 

Although at first sight these explanations for confinement are seemingly
different there are surprising relations between them which are not yet
understood. With the present level of understanding one has to note that
these theories are definitely not mutually exclusive but simply reveal only
different aspects of the confinement phenomenon.

In this talk I will focus on what one can learn from the infrared behaviour of
Landau gauge Greens functions about confinement.\footnote{See also the
recent review on this and related issues given in ref.\
\cite{Alkofer:2006jf}.}

\section{Infrared Structure of Landau gauge Yang-Mills theory}

\begin{figure}[th]
\centerline{\psfig{file=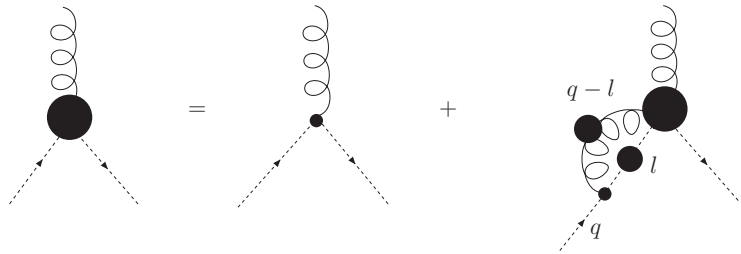,width=100mm}}
\vspace*{8pt}
\caption{The Dyson-Schwinger equation for the ghost-gluon vertex.
\label{GhGlDSE}}
\end{figure}

The starting point for our considerations is the Dyson-Schwinger equation for
the ghost-gluon vertex function as depicted in fig.\ \ref{GhGlDSE}. 
In the Landau gauge the gluon propagator 
is transverse, and thus one has the relation
\begin{equation}
l_\mu D_{\mu \nu}(l-q) = q_\mu D_{\mu \nu}(l-q) \, ,
\end{equation}
which immediately allows one to conclude that the ghost-gluon 
vertex stays finite when
the outgoing ghost momentum vanishes, {\it i.e.} when
$q_\mu \rightarrow 0$~\cite{Taylor:1971ff}. This argument is valid to all orders
in perturbation theory, a truely non-perturbative justification of the related
infrared finiteness has been given in refs.\
\cite{Lerche:2002ep,Cucchieri:2004sq,Schleifenbaum:2004id}.

\begin{figure}[th]
\centerline{\psfig{file=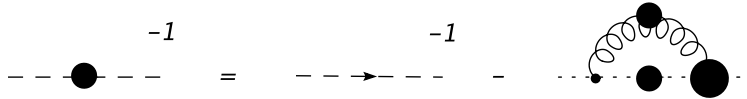,width=100mm}}
\vspace*{8pt}
\caption{The Dyson-Schwinger equation for the ghost propagator.
\label{GhDSE}}
\end{figure}

This property of the ghost-gluon vertex 
makes the Dyson-Schwinger equation for the ghost propagator,
see fig.\ \ref{GhDSE}, tractable.
The only unknowns in the deep infrared are the gluon and the ghost propagators.
In Landau gauge these propagators 
are parametrized by
two invariant functions, denoted here $Z(k^2)$ and $G(k^2)$, respectively.
In Euclidean momentum space one has
\begin{eqnarray} 
        D_{\mu\nu}(k) = \frac{Z(k^2)}{k^2} \, \left( \delta_{\mu\nu} -
        \frac{k_\mu k_\nu}{k^2} \right)  \; ,\quad
        D_G(k)  &=& - \frac{G(k^2)}{k^2}          \;.
        \nonumber \\
\end{eqnarray}

After renormalization the functions $Z(k^2)$ and $G(k^2)$ depend also on the 
renormalization  scale $\mu$. Furthermore, assuming that the QCD Green
functions can be expanded in asymptotic series, the integral in the ghost
Dyson--Schwinger equation can be split up in three pieces, an infrared
integral, an ultraviolet integral,  and an expression for the ghost wave
function renormalization. As a matter of fact, it is the resulting equation for 
the latter quantity which allows one to extract definite information
\cite{Watson:2001yv} without using any truncation or ansatz.

It turns out that the infrared behaviour of the gluon and ghost propagators
is given by power laws, and that the exponents are uniquely related such that
the gluon exponent is -2 times the ghost exponent
\cite{vonSmekal:1997is}. As we will see later on this
implies an infrared fixed point for the corresponding running coupling. The signs
of the exponents are such that the gluon propagator is infrared suppressed as
compared to the one for a free particle, the ghost propagator is infrared
enhanced.

\begin{figure}[th]
\centerline{\psfig{file=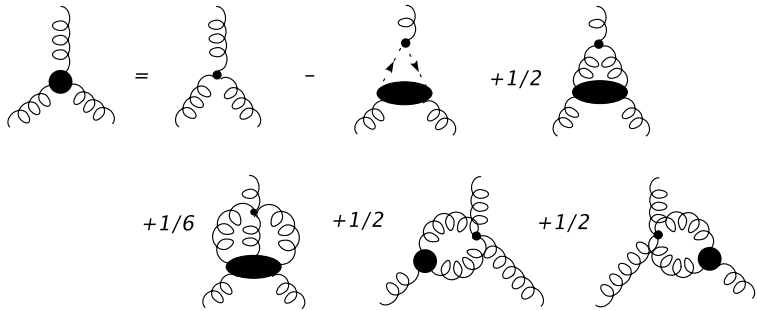,width=100mm}}
\vspace*{8pt}
\caption{The Dyson-Schwinger equation for the 3-gluon vertex.
\label{3GDSE}}
\end{figure}
\begin{figure}[th]
\centerline{\psfig{file=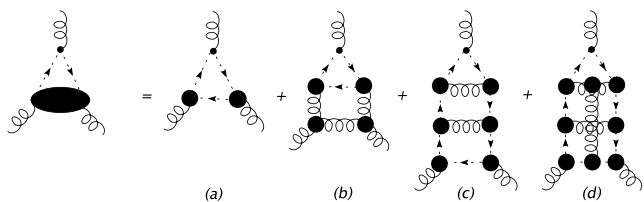,width=90mm}}
\vspace*{8pt}
\caption{An example for the skeleton expansion of the 
3-gluon vertex.
\label{3Gskel}}
\end{figure}
\begin{figure}[th]
\centerline{\psfig{file=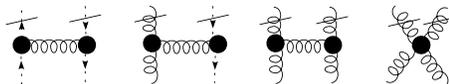,width=60mm}}
\vspace*{8pt}
\caption{Insertions to reconstruct higher orders in the skeleton expansion.
\label{3GskelIn}}
\end{figure}

The fact that the Yang-Mills propagators obey infrared power laws can be
employed to study the infrared behaviour of higher $n$-point functions. To this
end the corresponding $n$-point Dyson-Schwinger equations have been studied in a
skeleton expansion, {\it i.e.} a loop expansion using dressed propagators and
vertices. Furthermore, an asymptotic expansion has been applied to all primitively
divergent Green functions \cite{Alkofer:2004it}.
%\footnote{This scheme has been
%recently generalized to an arbitrary number of dimensions \cite{Huber:2007}.} 
As an example consider the Dyson-Schwinger equation for the 3-gluon vertex which
is diagrammatically represented in fig.\ \ref{3GDSE}.
Its skeleton expansion, see fig.\ \ref{3Gskel}, can be constructed 
via the insertions given in fig.\ \ref{3GskelIn}.
These insertions have vanishing infrared anomalous dimensions which implies
that the resulting higher order terms feature the same infrared scaling.
Based on this
the following general infrared behaviour for one-particle irreducible
Green functions with $2n$ external ghost legs and $m$ external
gluon legs can be derived\cite{Alkofer:2004it,Huber:2007}:
\begin{equation}
\Gamma^{n,m}(p^2) \sim (p^2)^{(n-m)\kappa + (1-n)(d/2-2)}  \label{IRsolution}
\end{equation}
where $\kappa$ is one yet undetermined parameter, and $d$ is the
space-time dimension.

\begin{figure}[th]
\centerline{\psfig{file=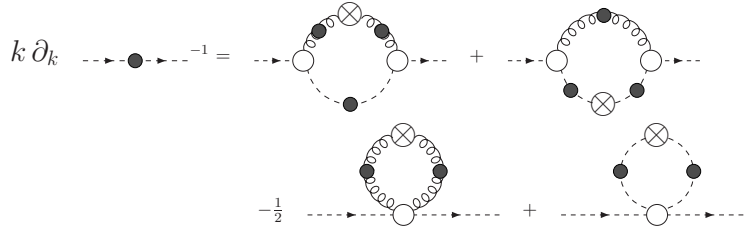,width=100mm}}
\vspace*{8pt}
\caption{The Exact Renormalization Group Equation
  for the ghost propagator.
\label{GhERGE}}
\end{figure}

Very recently it has been shown\cite{Fischer:2006vf} by exploiting
Dyson-Schwinger equations and Exact Renormalization Group Equations
({\it cf.} figs.\ \ref{GhDSE} and \ref{GhERGE} for the differences in these
equations for the ghost propagator)
that this infrared (IR) solution is unique.
It especially includes that
\begin{romanlist}[(b)]
\item
{the ghost propagator is IR divergent,}
\item
{the gluon propagator is IR suppressed,}
\item
{the ghost-gluon vertex is IR finite,}
\item
{the 3- and 4-gluon vertex are IR divergent,}
\item
{the ghost sector dominates the IR, and}
\item
{every coupling from an Yang-Mills vertex possesses an IR fixed point,\\
{\it i.e.} Infrared Yang-Mills theory is conformal.}
\end{romanlist}

\section{Yang-Mills running coupling: Infrared fixed point}

The infrared behaviour (\ref{IRsolution}) especially includes
\begin{align}
{ G(p^2) \sim (p^2)^{-\kappa}}\;,
&\quad { Z(p^2) \sim (p^2)^{2\kappa}}            \nonumber\\
{  \Gamma^{3g}(p^2) \sim (p^2)^{-3\kappa}} \;,
&\quad { \Gamma^{4g}(p^2) \sim (p^2)^{-4\kappa}} 
\end{align}
which allows to conclude that the running couplings as infered from these
verex functions possess an infrared fixed point:
\begin{eqnarray}
\displaystyle \alpha^{gh-gl}(p^2) &=&
\alpha_\mu \, { G^2(p^2)} \, { Z(p^2)}
\sim \frac{const_{gh-gl}}{N_c} ,
\end{eqnarray}
\begin{eqnarray}
\displaystyle \alpha^{3g}(p^2) &=&
\alpha_\mu \, { [\Gamma^{3g}(p^2)]^2} \, { Z^3(p^2)}
\sim \frac{const_{3g}}{N_c} ,
\end{eqnarray}
\begin{eqnarray}
\displaystyle \alpha^{4g}(p^2) &=&
\alpha_\mu \, {  \Gamma^{4g}(p^2)} \, { Z^2(p^2)}
\sim \frac{const_{4g}}{N_c} .
\end{eqnarray}

In particular, the infrared value of the coupling related to the ghost-gluon
vertex can be calculated\cite{Lerche:2002ep,Fischer:2002hn}:
\begin{equation}
\alpha^{gh-gl}(0)=\frac{4 \pi}{6N_c}
\frac{\Gamma(3-2\kappa)\Gamma(3+\kappa)\Gamma(1+\kappa)}{\Gamma^2(2-\kappa)
\Gamma(2\kappa)} 
\end{equation}
which yields $\alpha^{gh-gl}(0)=2.972$ for $N_c=3$ and $\kappa = (93 -
\sqrt{1201})/{98} \simeq 0.595353 $, which is the value obtained with a bare
ghost-gluon vertex.

\section{Positivity violation for the gluon propagator}

\begin{figure}[th]
\centerline{\psfig{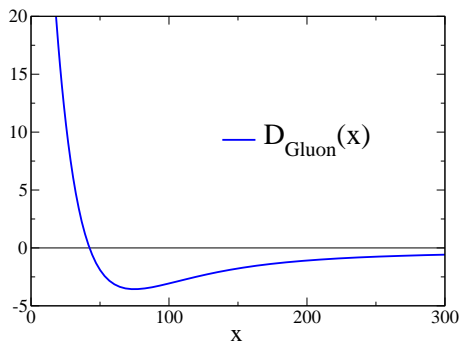}}
\vspace*{8pt}
\caption{The Fourier transform of the solution of the
 Dyson-Schwinger equation  for the gluon propagator, {\it cf.} ref.\
\protect \cite{Alkofer:2003jj}.
\label{Gluon-pos}}
\end{figure}
Positivity violation of the (space-time) propagator of transverse gluons has
been a long-standing conjecture for which there is now compelling evidence,
see  {\it e.g.\/} refs.~\cite{Alkofer:2003jj,Bowman:2007du} and references
therein. The basic feature is the infrared suppression of transverse gluons
caused by the infrared enhancement of ghost correlations. A simple argument
given by Zwanziger makes this obvious: An infrared vanishing gluon propagator
leads to a vanishing integral over the space-time gluon propagator, the latter 
being the Fourier transform of the momentum space gluon propagator. Therefore
one has
\begin{equation}
0=D_{gluon}(k^2=0) = \int d^4x \; \;\; \;  D_{gluon}(x)
\end{equation}
This implies that $D_{gluon}(x)$ has to be negative for some values of $x$.
And, as a matter of fact this behaviour is seen from fig.\ \ref{Gluon-pos} in
which the Fourier transform of the result for the gluon propagator is
displayed. As this behaviour clearly signals the confinement of tranverse gluons
\cite{vonSmekal:2000pz} it is certainly worth to have a closer look at the
underlying analytic structure of the gluon propagator.

To investigate the analytic structure of the gluon propagator
we parameterize the running coupling such that the numerical results
for Euclidean scales are quite accurately reproduced \cite{Fischer:2003rp}:
\begin{eqnarray}
\alpha_{\rm fit}(p^2) &=& \frac{\alpha_S(0)}{1+p^2/\Lambda^2_{\tt QCD}}\\
&+& \frac{4 \pi}{\beta_0} \frac{p^2}{\Lambda^2_{\tt QCD}+p^2}
\left(\frac{1}{\ln(p^2/\Lambda^2_{\tt QCD})}
- \frac{1}{p^2/\Lambda_{\tt QCD}^2 -1}\right) \nonumber
\end{eqnarray}
with $\beta_0=(11N_c-2N_f)/3$. In this expression the Landau pole has been
subtracted,
it is analytic in the complex $p^2$ plane except the real timelike axis
where the logarithm produces a cut for real $p^2<0$, and it obeys Cutkosky's 
rule.

The infrared exponent $\kappa$ is an irrational number, and thus  the
gluon propagator possesses a cut on the
negative real $p^2$ axis. It is possible to fit the solution for the gluon
propagator quite accurately without introducing further singularities
in the complex $p^2$ plane\cite{Alkofer:2003jj}:
\begin{equation}
Z_{\rm fit}(p^2) = w \left(\frac{p^2}{\Lambda^2_{\tt QCD}+p^2}\right)^{2 \kappa}
 \left( \alpha_{\rm fit}(p^2) \right)^{-\gamma} .
 \label{fitII}
\end{equation}
Hereby $w$ is a normalization parameter, and
$\gamma = (-13 N_c + 4 N_f)/(22 N_c - 4 N_f)$
is the one-loop value for
the anomalous dimension of the gluon propagator.
The discontinuity of (\ref{fitII}) along the cut
vanishes for $p^2\to 0^-$, diverges to $+\infty$ at $p^2=-\Lambda_{\tt QCD}^2$
and goes to zero for $p^2\to \infty$.

The function (\ref{fitII}) contains only four parameters:  
the overall magnitude
which due to renormalization properties is arbitrary (it
is determined via the choice of the renormalization scale),   the
scale $\Lambda_{\tt QCD}$,   the infrared exponent $\kappa$ and  the
anomalous dimension of the gluon $\gamma$. The latter two are not
free parameters: $\kappa$ is determined from the infrared properties,
 and for $\gamma$ the one-loop value is used. Thus one has found a
parameterization of the gluon propagator which has effectively only one
parameter, the scale $\Lambda_{\tt QCD}$.
It is important to note that the gluon propagator possesses a form such
that {\em Wick rotation is possible!}

Note that positivity violation for gluons is also found at very high
temperatures, even in the infinite temperature limit
\cite{Maas:2005hs,Cucchieri:2003di,Cucchieri:2007ta}.  For the gluons being
transverse to the medium it applies in both phases. This does not come as a
surprise: The infinite temperature limit corresponds to three-dimensional
Yang-Mills theory plus an additional Higgs-type field inherited from the $A_4$
field. The latter decouples in the infrared, the three-dimensional Yang-Mills
theory is as expected to be confining and thus the corresponding gluon modes
are positivity violating. This especially entails a solution of the Linde
problem \cite{Linde:1980ts}, see {\it e.g.} ref.\ \cite{Zwanziger:2006sc} and
references therein. Based on this scheme it is safe to conclude that the 
static chromomagnetic sector is  never deconfined, {\it cf.} ref.\ 
\cite{Arnold}.

\section{Ghost and Glue in a box}

\begin{figure}[bh]
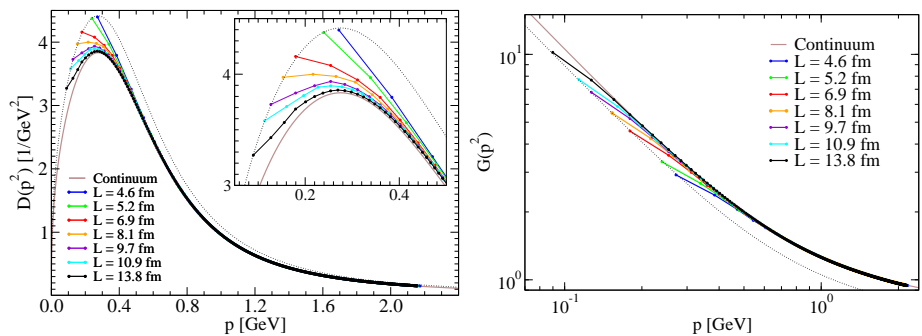

\centerline{\psfig{file=FinVolGl.eps,width=60mm}
\psfig{file=FinVolGh.eps,width=60mm}}
\vspace*{8pt}
\caption{The solution of the
 Dyson-Schwinger equation  for the gluon and the ghost propagators 
 at finite volumes (adapted from ref.\ \protect \cite{Fischer:2007pf}).
\label{Gluon-vol}}
\end{figure}

As can be noted from refs.\ \cite{Fischer:2002hn,Fischer:2003rp} the numerical
results for the ghost and gluon propagators compare very well to corresponding
recent lattice data. However, the values of the infrared exponents extracted
from lattice calculations do neither agree with the analytical obtained 
continuum results nor do they agree when compared against each other. A
comparison to lattice calculations in three and two spacetime dimensions
suggest that current lattice volumes are much too small for reliable extraction
of infrared exponents  \cite{Maas:2006qw,Maas:2007uv}.

\begin{figure}[th]
\centerline{\psfig{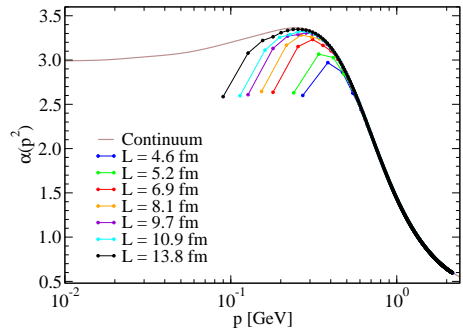}}
\vspace*{8pt}
\caption{The resulting running coupling at finite volumes (adapted from ref.\
\protect \cite{Fischer:2007pf}).
\label{alpha-vol}}
\end{figure}

At this point it is interesting to note that the Dyson-Schwinger equations can
be solved on a compact manifold with finite volume
\cite{Fischer:2005ui,Fischer:2007pf}. 
As can be seen from figs.\ \ref{Gluon-vol}
and \ref{alpha-vol} the approach to the continuum limit is quite slow.

\begin{figure}[bh]
\centerline{\psfig{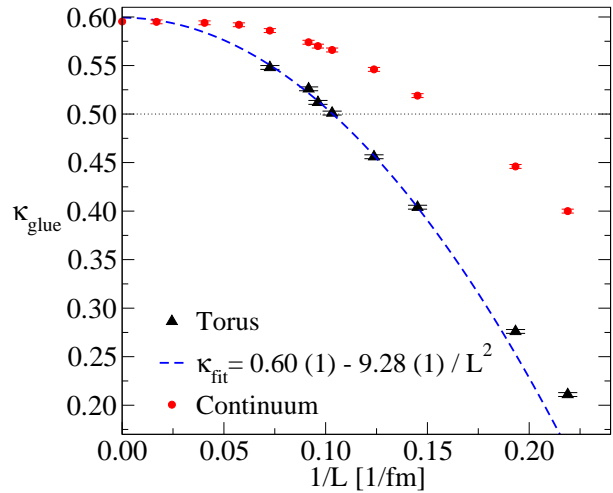}}
\vspace*{8pt}
\caption{The extracted value of the infrared exponent $\kappa$ as function of
the inverse of the torus length (adapted from ref.\
\protect \cite{Fischer:2007pf}).
\label{vol}}
\end{figure}
This slow approach can be understood from fig.\ \ref{vol}. The triangles
represent the torus results, the circles are the values of the infrared exponent
$\kappa$ as extracted from the continuum result when restricted to momenta
$p\le 2\pi/L$. This makes plain that for a precise extraction of the infrared
exponent a large separation of scales is needed. This, and not the existence of
a very small momentum scale inherent to the problem, is the reason why large
volumes are needed for a precise determination of the infrared exponent.

\begin{figure}[th]
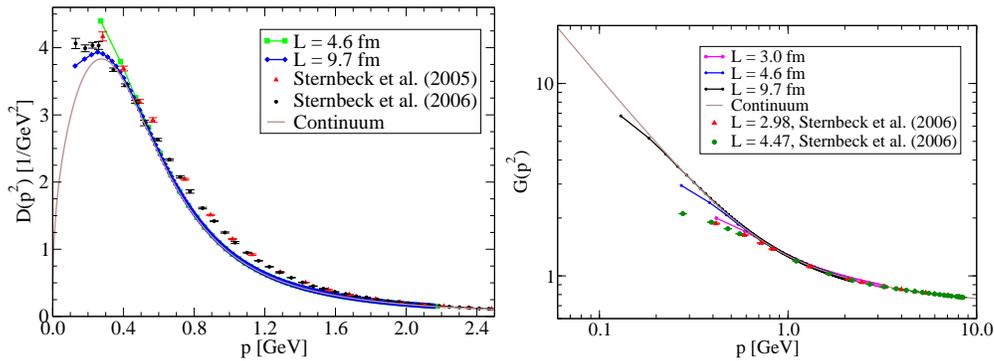

\centerline{\psfig{file=FinVolGlL.eps,width=65mm}
\psfig{file=FinVolGhL.eps,width=65mm}}
\vspace*{8pt}
\caption{The solution of the
 Dyson-Schwinger equations  for the gluon and ghost propagators 
 at finite volumes
 compared to lattice data (adapted from ref.\
\protect \cite{Fischer:2007pf}).
\label{Gluon-lat}}
\end{figure}

In fig.\ \ref{Gluon-lat} the results of the
Dyson-Schwinger equations for the propagators are compared to recent lattice
data\cite{Sternbeck:2006cg} using very similar volumes. The agreement is, at
least, satisfactory.

\section{Dynamically induced scalar quark confinement}

\begin{figure}[th]
\centerline{\psfig{file=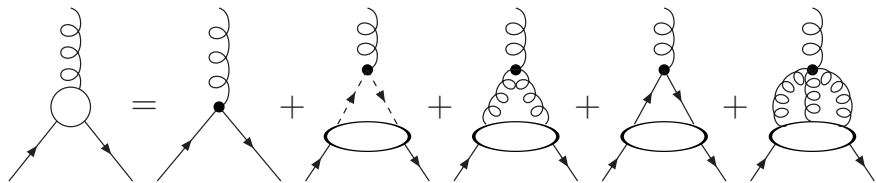,width=120mm}}
\vspace*{8pt}
\caption{The 
 Dyson-Schwinger equation  for the quark-gluon vertex.
\label{QGV}}
\end{figure}

\begin{figure}[th]
\centerline{\psfig{file=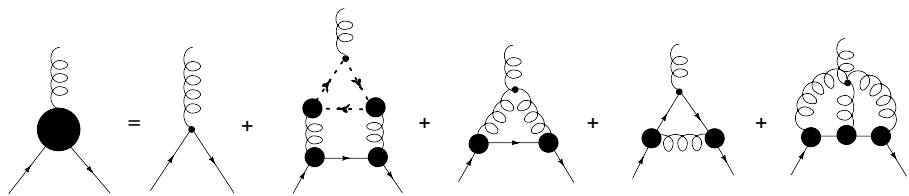,width=120mm}}
\vspace*{8pt}
\caption{The skeleton expansion   for the quark-gluon vertex.
\label{QGV-skel}}
\end{figure}

From the presentation above it is quite obvious how gluon confinement  works in
a covariant gauge. However, given the infrared suppression of the gluon
propagator quark confinement is even seemingly more mysterious than ever.  To
proceed in the same spirit as above one studies the Dyson-Schwinger equation
for the quark propagator. It turns out that the precise structure of the  quark
propagator depends crucially on the quark-gluon
vertex\cite{Roberts:1994dr,Alkofer:2000wg,Fischer:2006ub,Fischer:2003rp}.
Therefore a detailed study of this three-point function, and especially its 
infrared behaviour, is mandatory. Its Dyson-Schwinger equation is
diagrammatically depicted in fig.\ \ref{QGV}, its skeleton expansion in
fig.\ \ref{QGV-skel}.

At this point one has to notice a drastic difference of the quarks as compared
to Yang-Mills fields: They possess a current, {\it i.e.} a tree-level,
mass.\footnote{Even if this were not the case one expects dynamical chiral
symmetry breaking and thus dynamical mass generation to occur.} To extend the
infrared analysis of Yang-Mills theory described above to full QCD
\cite{Alkofer:2006gz} we thus want to concentrate first on the quark sector of
quenched QCD and choose the masses of the valence quarks to  be large, {\it
i.e.\/}  $m > \Lambda_{\tt QCD}$.   The remaining scales below $\Lambda_{\tt
QCD}$ are those of the external momenta of the propagators and vertex functions. 
The relevant infrared limit is the one where all these external momenta approch
zero. Then the Dyson-Schwinger equations can be used  
to determine the selfconsistent solutions  in
terms of powers of the small external momentum scale $p^2 \ll \Lambda_{\tt
QCD}$. The equations which have to be considered in addition to the ones of
Yang-Mills theory are the one for the quark propagator and the quark-gluon
vertex.

The dressed quark-gluon vertex $\Gamma_\mu$ consists in general of twelve
linearly independent Dirac tensors. Some of those would be forbidden if chiral
symmetry would be realized in the Wigner-Weyl mode. On the other hand, these
tensor structures can be non-vanishing either if chiral symmetry is explicitely
broken by current masses and/or chiral symmetry is realized in Nambu-Goldstone
mode ({\it i.e.} spontaneously broken). From a solution of the Dyson-Schwinger
equations we infer that  these ``Dirac-scalar'' structures are, in the chiral
limit, generated non-perturbatively together with the dynamical quark mass
function in a self-consistent fashion: Dynamical chiral symmetry breaking
reflects itself not only in the propagator but also in the quark-gluon vertex
function.

An infrared analysis of the full set of Dyson-Schwinger equations reveals 
an infrared divergent solution for the quark-gluon vertex.
Hereby,  Dirac vector and {\em ``scalar''}
components of this  vertex are infrared divergent with exponent $-\kappa -
\frac 1 2$ \cite{Alkofer:2006gz}. A numerical solution of a truncated set of
Dyson-Schwinger equations confirms this infrared behavior. Again, 
the diagrams containing  ghost loops dominate.  Thus all
infrared effects from the Yang-Mills sector are generated by the infrared
asymptotic theory described above. More importantly, in the quark sector the
driving pieces of this solution are the scalar Dirac amplitudes of the
quark-gluon vertex and the scalar part of the quark propagator. Both pieces are
only present when chiral symmetry is broken, either explicitely or dynamically.

For the coupling related to the quark-gluon vertex
we obtain, using
\begin{equation}
{\Gamma^{qg}(p^2) \sim (p^2)^{-1/2-\kappa}} \,, \ \
{ Z_f(p^2) \sim const}\,, \ \ { Z(p^2) \sim (p^2)^{2\kappa}} ,
\end{equation}
that 
\begin{equation}
\alpha^{qg}(p^2) = \alpha_\mu \,
{ [\Gamma^{qg}(p^2)]^2} \, { [Z_f(p^2)]^2}\,
{ Z(p^2)} \sim  \frac{const_{qg}}{N_c} \frac{1}{p^2} ,
\end{equation}
{\it i.e.} that it is  singular in the infrared contrary to the
couplings  from the Yang-Mills vertices.

With similar methods one finds for the four-quark function  an anomalous
infrared exponent ${-2}$.  Note that the static quark potential can be obtained 
from this four-quark one-particle irreducible Greens function, which, including
the canonical dimensions, behaves like $(p^2)^{-2}$for $p^2\to0$.
Therefore employing the well-known relation for a function $F\propto (p^2)^{-2}$
one obtains
\begin{equation}
V({\bf r}) = \int \frac{d^3p}{(2\pi)^3}  F(p^0=0,{\bf p})  e^{i {\bf p r}}
\ \ \sim \ \ |{\bf r} |
\end{equation}
for the static quark-antiquark potential $V({\bf r})$. 
We conclude at this point that, given the infrared divergence of the
quark-gluon vertex as found in the solution of the coupled system of
Dyson-Schwinger equations, the vertex overcompensates the infared suppression 
of the gluon propagator, and one therefore obtains a linear rising potential.
In addition, this potential is dynamically induced and dominantly
scalar.

However, there are two caveats. First, the uniqueness of this solution could
not be shown. Second, because most of the terms in the skeleton expansion 
of the four-quark function are equally enhanced in the infrared 
the string tension could only be calculated by summing over an
infinite number of diagrams. This property alleviates the usefulness
of the approach but it had to be expected in the first place. Since already an
effective, nonperturbative one-gluon exchange generates the confining potential
one is confronted with the problem of unwanted van-der-Waals forces. To avoid
such forces there has to occur a precise cancelation of them amongst the 
infinitely many terms contributing to the long-range part of the potential.

An interesting limit can, however, be studied. Suppose chiral symmetry is
artificially kept in Wigner-Weyl mode, {\it i.e.} in the chiral limit we force
the quark mass term as well as the ``scalar'' terms in the quark-gluon vertex
to be zero. We then find $-\kappa$ as infrared exponent for the vertex, and
the  resulting running coupling from the quark-gluon vertex is no longer
diverging but goes to a fixed point in the infrared similar to the couplings
from the Yang-Mills vertices. Correspondingly, one obtains a constant for 
$\Gamma^{0,0,2}(p^2=0)$ and 
$$
V({\bf r}) \sim  \int \frac
{d^3p}{(2\pi)^3}\frac{1}{p^2} e^{i {\bf p r}}
 \ \ \sim \ \ {\bf\frac{1}{|r|}} .
$$
The ``forced'' restoration of chiral symmetry is therefore directly linked with
the disappearance of  quark confinement. The infared properties of the
quark-gluon vertex in the ``unforced'' solution thus constitute a novel
mechanism that  directly links chiral symmetry breaking with confinement.

\section{Summary}

To summarize the most important findings as infered from the infrared analysis
of all one-particle irreducible Green functions of Landau gauge QCD we note:

\smallskip

\noindent 
Gluons are confined by ghosts,  and positivity of transverse gluons is
violated.

\smallskip

\noindent 
The analytic structure of the resulting gluon propagator is 
such that Wick rotation is possible.

\smallskip

\noindent 
In the Yang-Mills sector the strong running coupling is infrared finite
whereas the running coupling from the quark-gluon vertex is infrared divergent.

\smallskip

\noindent 
Chiral symmetry is dynamically broken, and this takes place
in the quark propagator and
the quark-gluon vertex.

\smallskip

\noindent 
  We have provided evidence that static quark confinement in the Landau 
  gauge is
  due to the infrared divergence of the quark-gluon vertex. \\
  In the infrared this vertex is dominated by its scalar components
  thereby inducing a
  relation between confinement and
  broken chiral symmetry.

\section*{Acknowledgements}

RA thanks the organizers of {\em X Hadron Physics 2007}
for inviting him to give a seminar at this extraordinarily 
interesting workshop.

\noindent
This work has been supported in part by the DFG under contract
AL 279/5-1 and by the FWF under contract M979-N16.


\begin{thebibliography}{0}

\bibitem{Alkofer:2006fu}
  R.~Alkofer and J.~Greensite,
  %``Quark Confinement: The Hard Problem of Hadron Physics,''
  J.\ Phys.\ {\bf G34} (2007) S3
  [arXiv:hep-ph/0610365].
  %%CITATION = HEP-PH/0610365;%%

\bibitem{Mandelstam}
  S.~Mandelstam,
  %``Vortices And Quark Confinement In Nonabelian Gauge Theories,''
  Phys.\ Rept.\  {\bf 23} (1976) 245.
  %%CITATION = PRPLC,23,245;%%

\bibitem{Pisa}
   A.~Di Giacomo, B.~Lucini, L.~Montesi and G.~Paffuti,
  %``Colour confinement and dual superconductivity of the vacuum. I,''
  Phys.\ Rev.\ D {\bf 61}  (2000) 034503
  [arXiv:hep-lat/9906024];
  %%CITATION = HEP-LAT 9906024;%%
%  A.~Di Giacomo, B.~Lucini, L.~Montesi and G.~Paffuti,
  %``Colour confinement and dual superconductivity of the vacuum. II,''
  Phys.\ Rev.\ D {\bf 61} (2000) 034504
  [arXiv:hep-lat/9906025].
  %%CITATION = HEP-LAT 9906025;%%

\bibitem{Greensite} J.~Greensite,
  Prog.\ Part.\ Nucl.\ Phys.\  {\bf 51} (2003) 1
  [arXiv:hep-lat/0301023].
%%CITATION = HEP-LAT 0301023;%%

\bibitem{Gribov} V.\ Gribov, Nucl. Phys. B\textbf{139} (1978) 1.
%%CITATION = NUPHA,B139,1;%%
\bibitem{Dan1} D. Zwanziger, Nucl. Phys. B\textbf{518} (1998) 237;
 Phys. Rev. Lett. \textbf{90} (2003) 102001 
[arXiv: hep-th/0209105].

\bibitem{Alkofer:2000wg}
  R.~Alkofer and L.~von Smekal,
  % ``The infrared behavior of QCD Green's functions: Confinement, dynamical
  %symmetry breaking, and hadrons as relativistic bound states,''
  Phys.\ Rept.\  {\bf 353}  (2001) 281
  [arXiv:hep-ph/0007355].
  %%CITATION = HEP-PH 0007355;%%

\bibitem{Fischer:2006ub}
  C.~S. Fischer, J. Phys. G: Nucl. Part. Phys. {\bf 32} (2006) R253
  [arXiv:hep-ph/0605173].
  %%CITATION = HEP-PH 0605173;%%

\bibitem{Aguilar:2004sw}
  A.~C.~Aguilar and A.~A.~Natale,
  %``A dynamical gluon mass solution in a coupled system of the  Schwinger-Dyson
  %equations,''
  JHEP {\bf 0408} (2004) 057
  [arXiv:hep-ph/0408254].
  %%CITATION = JHEPA,0408,057;%%
  
\bibitem{Maldacena:1998im}
  J.~M.~Maldacena,
  %``Wilson loops in large N field theories,''
  Phys.\ Rev.\ Lett.\  {\bf 80} (1998) 4859
  [arXiv:hep-th/9803002]. 
  %%CITATION = HEP-TH 9803002;%%

\bibitem{Alkofer:2006jf}
  R.~Alkofer,
  %``QCD Green Functions and their Application to Hadron Physics,''
  Brazilian J.\ Phys.\ {\bf 37} (2007) 144
  [arXiv:hep-ph/0611090].
  %%CITATION = HEP-PH/0611090;%%

\bibitem{Taylor:1971ff}
J.~C. Taylor, {Nucl. Phys.} {\bf B33} (1971) 436.
%%CITATION = NUPHA,B33,436;%%.

\bibitem{Lerche:2002ep}
C.~Lerche and L.~von Smekal,
%``On the infrared exponent for gluon and ghost propagation in Landau  gauge QCD,''
Phys.\ Rev.\ D {\bf 65}  (2002) 125006
[arXiv:hep-ph/0202194].
 %%CITATION = HEP-PH 0202194;%%

\bibitem{Cucchieri:2004sq}
A.~Cucchieri, T.~Mendes, and A.~Mihara {JHEP} {\bf 12} (2004) 012
[arXiv:hep-lat/0408034].
%%CITATION = HEP-LAT 0408034;%%.

\bibitem{Schleifenbaum:2004id}
W.~Schleifenbaum {\it et al.},
%W.~Schleifenbaum, A.~Maas, J.~Wambach, and R.~Alkofer,
{Phys. Rev.} {\bf   D72} (2005) 014017 
[arXiv:hep-ph/0411052].
%%CITATION = HEP-PH 0411052;%%.

\bibitem{Watson:2001yv}
  P.~Watson and R.~Alkofer,
  %``Verifying the Kugo-Ojima confinement criterion in Landau gauge QCD,''
  Phys.\ Rev.\ Lett.\  {\bf 86} (2001) 5239 
  [arXiv:hep-ph/0102332].
  %%CITATION = HEP-PH 0102332;%%

\bibitem{vonSmekal:1997is}
L.~von Smekal, R.~Alkofer and A.~Hauck,
%``The infrared behavior of gluon and ghost propagators in Landau gauge  QCD,''
Phys.\ Rev.\ Lett.\  {\bf 79}, 3591 (1997)
[arXiv:hep-ph/9705242];
%%CITATION = HEP-PH 9705242;%%
%\bibitem{vonSmekal:1998is}
L.~von Smekal, A. Hauck and R.~Alkofer,
%``A solution to coupled Dyson-Schwinger equations for gluons and ghosts
%in Landau gauge,''
Annals Phys.\  {\bf 267}, 1 (1998)
%[Erratum-ibid.\  {\bf 269}, 182 (1998)]
[arXiv:hep-ph/9707327];
%%CITATION = HEP-PH 9707327;%%
%\cite{Hauck:1998fz}
%\bibitem{Hauck:1998fz}
A.~Hauck, L.~von Smekal and R.~Alkofer,
%``Solving a coupled set of truncated {QCD} Dyson-Schwinger equations,''
Comput.\ Phys.\ Commun.\  {\bf 112} (1998) 166
[arXiv:hep-ph/9804376].
%%CITATION = HEP-PH 9804376;%%

\bibitem{Alkofer:2004it}
  R.~Alkofer, C.~S.~Fischer and F.~J.~Llanes-Estrada,
  % ``Vertex functions and infrared fixed point in Landau gauge SU(N)  Yang-Mills
  %theory,''
  Phys.\ Lett.\ B {\bf 611} (2005)  279 
  [arXiv:hep-th/0412330];
  %%CITATION = HEP-TH 0412330;%%
%\cite{Alkofer:2006xz}
%\bibitem{Alkofer:2006xz}
%  R.~Alkofer, P.~Bicudo, S.~R.~Cotanch, C.~S.~Fischer and F.~J.~Llanes-Estrada,
  R.~Alkofer {\it et al.},
  %``Soft gluons are heavy and rowdy,''
  arXiv:nucl-th/0601032.
  %%CITATION = NUCL-TH/0601032;%%

\bibitem{Huber:2007} M.~Huber, R.~Alkofer, C.~S.~Fischer and K.~Schwenzer,
arXiv:0705.3809 [hep-ph].
  %%CITATION = ARXIV:0705.3809;%%.

\bibitem{Fischer:2006vf}
  C.~S.~Fischer and J.~M.~Pawlowski,
  %``Uniqueness of infrared asymptotics in Landau gauge Yang-Mills theory,''
  Phys.\ Rev.\  D {\bf 75} (2007) 025012
  [arXiv:hep-th/0609009].
  %%CITATION = PHRVA,D75,025012;%%

\bibitem{Fischer:2002hn}
C.~S.~Fischer  and R.~Alkofer,
%{\it Infrared Exponents and Running Coupling of SU(N) Yang--Mills Theories},
Phys. Lett. {\bf B536}  (2002) 177
[arXiv:hep-ph/0202202];
%%CITATION = HEP-PH 0202202;%%
C.~S.~Fischer, R.~Alkofer and H.~Reinhardt,
%{\it The Elusiveness of Infrared Critical Exponents in Landau
%Gauge Yang-Mills Theories},
Phys.\ Rev.\ {\bf D65} {2002} 094008  [arXiv:hep-ph/0202195];
%%CITATION = HEP-PH 0202195;%%
%\bibitem{Alkofer:2002ne}
R.~Alkofer, C.~S.~Fischer and L.~von Smekal,
%``The infrared behaviour of the running coupling in Landau gauge QCD,''
Acta Phys.\ Slov.\  {\bf 52}(2002) 191  [arXiv:hep-ph/0205125].
%%CITATION = HEP-PH 0205125;%%

\bibitem{Alkofer:2003jj}
R.~Alkofer {\it et al.},
  %R.~Alkofer, W.~Detmold, C.~S.~Fischer and P.~Maris,
  %``Analytic properties of the Landau gauge gluon and quark propagators,''
  Phys.\ Rev.\ D {\bf 70} (2004) 014014
  [arXiv:hep-ph/0309077];
  %%CITATION = HEP-PH 0309077;%%
 %\cite{Alkofer:2003jk}
 %\bibitem{Alkofer:2003jk}
 %  R.~Alkofer, W.~Detmold, C.~S.~Fischer and P.~Maris,
 %   ``Analytic structure of the gluon and quark propagators in Landau gauge
  %QCD,''
  Nucl.\ Phys.\ Proc.\ Suppl.\  {\bf 141} (2005) 122.
  %[arXiv:hep-ph/0309078].
  %%CITATION = HEP-PH 0309078;%%

\bibitem{Bowman:2007du}
  P.~O.~Bowman {\it et al.},
  %``Scaling behavior and positivity violation of the gluon propagator in full
  %QCD,''
  arXiv:hep-lat/0703022.
  %%CITATION = HEP-LAT/0703022;%%

\bibitem{vonSmekal:2000pz}
  L.~von Smekal and R.~Alkofer,
  %``What the infrared behavior of QCD Green functions can tell us about
  %confinement in the covariant gauge,''
  arXiv:hep-ph/0009219.
  %%CITATION = HEP-PH/0009219;%%

\bibitem{Fischer:2003rp}
  C.~S.~Fischer and R.~Alkofer,
  % ``Non-perturbative propagators, running coupling and dynamical quark mass of
  %Landau gauge QCD,''
  Phys.\ Rev.\ D {\bf 67}, 094020 (2003)
  [arXiv:hep-ph/0301094].
  %%CITATION = HEP-PH 0301094;%%


\bibitem{Maas:2005hs}
  A.~Maas, J.~Wambach and R.~Alkofer,
  %``The high-temperature phase of Landau-gauge Yang-Mills theory,''
  Eur.\ Phys.\ J.\ C{\bf 42}, 93 (2005)
  [arXiv:hep-ph/0504019];
  %%CITATION = HEP-PH 0504019;%%
%\bibitem{Maas:2004se}
A.~Maas {\it et al.},
  %A.~Maas, J.~Wambach, B.~Gruter and R.~Alkofer,
  %``High-temperature limit of Landau-gauge Yang-Mills theory,''
  Eur.\ Phys.\ J.\ C{\bf 37}, 335 (2004)
  [arXiv:hep-ph/0408074];
  %%CITATION = HEP-PH 0408074;%%

\bibitem{Cucchieri:2003di}
  A.~Cucchieri, T.~Mendes and A.~R.~Taurines,
  %``SU(2) Landau gluon propagator on a 140**3 lattice,''
  Phys.\ Rev.\ D {\bf 67}, 091502 (2003)
  [arXiv:hep-lat/0302022].
  %%CITATION = HEP-LAT 0302022;%%

\bibitem{Cucchieri:2007ta}
  A.~Cucchieri, A.~Maas and T.~Mendes,
  %``Infrared properties of propagators in Landau-gauge pure Yang-Mills   theory
  %at finite temperature,''
  Phys.\ Rev.\  D {\bf 75} (2007) 076003
  [arXiv:hep-lat/0702022].
  %%CITATION = PHRVA,D75,076003;%%

\bibitem{Linde:1980ts}
  A.~D.~Linde,
  %``Infrared Problem In Thermodynamics Of The Yang-Mills Gas,''
  Phys.\ Lett.\  B {\bf 96} (1980) 289.
  %%CITATION = PHLTA,B96,289;%%

\bibitem{Zwanziger:2006sc}
  D.~Zwanziger,
  %``Equation of state of gluon plasma from local action,''
  arXiv:hep-ph/0610021.
  %%CITATION = HEP-PH/0610021;%%

\bibitem{Arnold} P.\ Arnold, these proceedings.

\bibitem{Maas:2006qw}
  A.~Maas, A.~Cucchieri and T.~Mendes,
  %``On the infrared behavior of Green's functions in Yang-Mills theory,''
  Braz.\ J.\ Phys.\  {\bf 37N1B} (2007) 219
  [arXiv:hep-lat/0610006].
  %%CITATION = BJPHE,37N1B,219;%%

\bibitem{Maas:2007uv}
  A.~Maas,
  %``Two- and three-point Green's functions in two-dimensional Landau-gauge
  %Yang-Mills theory,''
  arXiv:0704.0722 [hep-lat].
  %%CITATION = ARXIV:0704.0722;%%

\bibitem{Fischer:2005ui}
  C.~S.~Fischer, B.~Gruter and R.~Alkofer,
  %``Solving coupled Dyson-Schwinger equations on a compact manifold,''
  Annals Phys.\  {\bf 321}, 1918 (2006)
  [arXiv:hep-ph/0506053].
  %%CITATION = HEP-PH 0506053;%%

\bibitem{Fischer:2007pf}
  C.~S.~Fischer, A.~Maas, J.~M.~Pawlowski and L.~von Smekal,
  %``Large volume behaviour of Yang-Mills propagators,''
   Annals Phys. (2007), in print [arXiv:hep-ph/0701050].
  %%CITATION = HEP-PH/0701050;%%

\bibitem{Sternbeck:2006cg}
  A.~Sternbeck {\it et al.},
  %A.~Sternbeck, E.~M.~Ilgenfritz, M.~Muller-Preussker, A.~Schiller and I.~L.~Bogolubsky,
  %``Lattice study of the infrared behavior of QCD Green's functions in Landau
  %gauge,''
  PoS {\bf LAT2006} (2006) 076
  [arXiv:hep-lat/0610053];
  %%CITATION = POSCI,LAT2006,076;%%
  A.~Sternbeck,
  %``The infrared behavior of lattice QCD Green's functions,''
  PhD thesis [arXiv:hep-lat/0609016].
  %%CITATION = HEP-LAT/0609016;%%

\bibitem{Roberts:1994dr}
  C.~D.~Roberts and A.~G.~Williams,
  %``Dyson-Schwinger equations and their application to hadronic physics,''
  Prog.\ Part.\ Nucl.\ Phys.\  {\bf 33} (1994) 477
  [arXiv:hep-ph/9403224].
  %%CITATION = PPNPD,33,477;%%
  
\bibitem{Alkofer:2006gz}
  R.~Alkofer, C.~S.~Fischer and F.~J.~Llanes-Estrada,
  %``Dynamically induced scalar quark confinement,''
  arXiv:hep-ph/0607293.
  %%CITATION = HEP-PH 0607293;%%
  


\end{thebibliography}
\end{document}